# Electrical Conductivity of Collapsed Multilayer Graphene Tubes

**D. Mendoza**

Instituto de Investigaciones en Materiales, Universidad Nacional Autónoma de México, Coyoacán D. F. 04510, México
Email: doroteo@unam.mx

**Abstract**

Synthesis of multilayer graphene on copper wires by a chemical vapor deposition method is reported. After copper etching, the multilayer tube collapses forming stripes of graphitic films, their electrical conductance as a function of temperature indicate a semiconductor-like behavior. Using the multilayer graphene stripes, a cross junction is built and owing to its electrical behavior we propose that a tunneling process exists in the device.

**Keywords:** Multilayer Graphene; Electrical Conductivity; Klein Tunneling; Chemical Vapor Deposition

## 1. Introduction

Graphene and related systems are of great interest due to their physical properties and also by their possible applications. Although graphene is a zero band gap semimetal, chemical and geometrical modifications of this material allow different electronic and optical properties. For example, in hydrogenated [1] and fluorinated [2] graphene a band gap is opened, besides graphene nanoribbons can also have a gap and other novel electronic properties [3]. Nanoholes in periodic arrangements can induce magnetic behavior [4] or band gap opening [5]. Change of the local electrical conductivity by moving the Fermi level, using an external bias for example, following predetermined geometrical patterns has been proposed as a mean of controlling the propagation of electromagnetic modes. That is, graphene has been proposed as a metamaterial with cloaking properties [6] or more complex functionalities based on transformation optics [7].

On the other hand, under radial deformation carbon nanotube eventually collapses and for radius greater than a crossover value, the collapsed state is more stable than the tubular geometry [8]. In some cases collapse induces metallic carbon nanotubes to become semiconducting, and vice versa [9]. In this work, we explore the synthesis and the study of some electrical properties of similar collapsed tubes, but at the macroscopic scale, depositing multilayer graphene on copper wires by a chemical vapor deposition (CVD) method.

CVD method is a promising technique to grow graphene in large area using copper foil as a catalyst and hydrocarbon vapors as the carbon source [10]. Due to the low carbon solubility in copper, the reaction of hydrocarbon species is limited to a region near to the copper surface, allowing the synthesis of graphene in almost any arbitrary form of the copper surface. Here we exploit this advantage to synthesize few-layer graphene [11] on cylindrical copper surface by CVD at atmospheric pressure.

## 2. Experimental Details

Multilayer graphene were grown on copper wire by a chemical vapor deposition method at ambient pressure. The wires were heated in an hydrogen ambient with 25 sccm flux up to 1000 $^{O}$C and maintained at this temperature by 15 min to anneal the copper wire. After this process hydrogen flux were adjusted to 90 sccm and methane was added with 25 sccm flux during 15 minutes, at the end of the process, methane flux was cut off and the furnace was turned off and the sample was cooled in the hydrogen atmosphere to room temperature. Copper was etched in a ferric nitrate aqueous solution, washed with deionized water and the carbonaceous film was transferred to copper grids for transmission electron microscopy (TEM) observation, and to glass substrates for optical observation and electrical characterization. Care should be taken because in this process the films may be damaged by tearing or be folded in some regions. The electrical characterization was carried out in a chamber equipped with a heater and a cold stage for cooling below room temperature using liquid nitrogen. Silver strips obtained by thermal evaporation 1 mm apart were used as electrical contacts, at a fixed bias voltage the electrical current through the sample was measured under vacuum conditions (~$10^{-4}$ Torr).

## 3. Results and Discussion

### 3.1. Collapsed Multilayer Graphene Tubes

Two kinds of wires with different diameters were used, the diameter being measured after the synthesis process using an optical microscope such as is observed in **Figure 1(a)**. The average diameter of the thinner one is 63μm and 163μm for the thicker wire (at this level, the thickness of the multilayer graphene is negligible). In **Figure1(b)** the collapsed multilayer graphene tube on

**(a)**

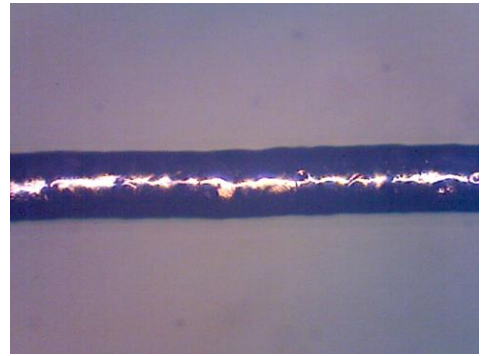

**(b)**

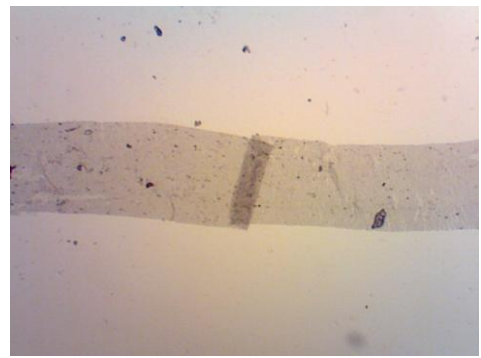

**Figure 1.** (a) Copper wire with average diameter of 63μm with bilayer graphene on its surface. (b) Collapsed tube of multilayer graphene after copper etching with an average width of 92μm transferred to a glass substrate, the darker fringe on the middle of this image is a folded region of the film.

glass substrate as observed by an optical microscope in the transmission mode is shown.

The mean width (W) of the collapsed tubes is 92μm and 203μm for the thinner and thicker copper wires, respectively. It should be noted that similar structures have been reported by other authors but using nickel nanowires as the template [12].

**(a)**

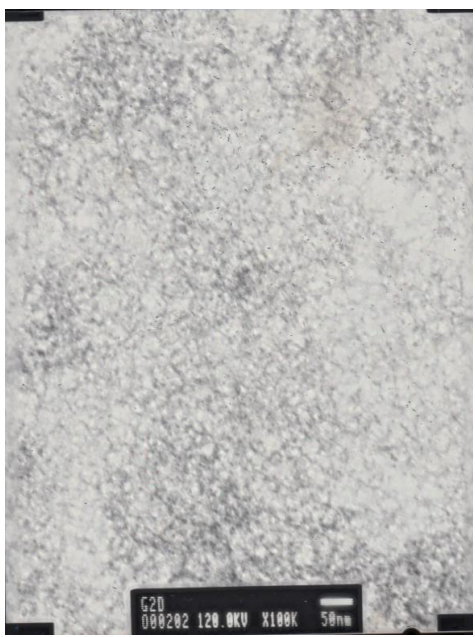

**(b)**

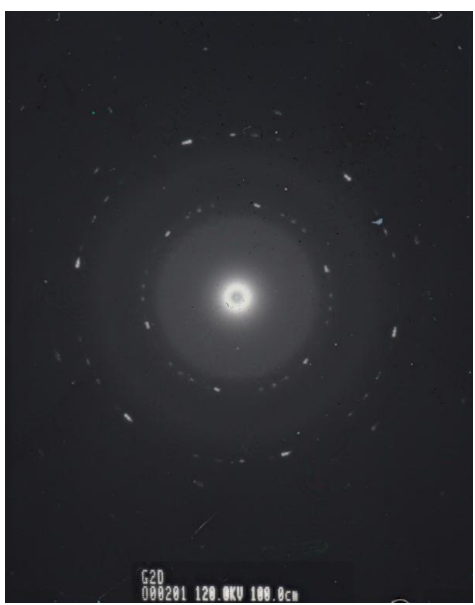

**Figure 2.** (a) TEM image of freely supported multilayer film, scale bar is 50 nm. (b) Diffraction pattern of the same film.

Using a digital image obtained through the optical microscope to measure the transmittance of the film [13], we found a value of ~91%, and using the generalized rule of 2.3% of absorbance per graphene layer [14], then our films approximately have 4 layers of graphene. But the film is formed by the collapsed tube, therefore this means that under our experimental conditions of synthesis, mainly a bilayer graphene is formed on the surface of the copper wire.

As complementary information, in **Figure 2** TEM image (a) and the diffraction pattern (b) of the freely supported collapsed tube are shown. Note from **Figure 2(a)** that the films are somehow inhomogeneous, and the diffraction pattern shows that the sample is polycrystalline.

### 3.2. Electrical Characterization of Collapsed Tube Stripes

In **Figure 3** electrical conductance as a function of temperature for the two kinds of samples is presented. A linear behavior for both samples appears to be a good description of the electrical conductivity, as is observed by the straight line fit to the upper curve in **Figure 3(b)**. Using the results presented in **Figure 3(b)** at T=300K, and the following geometrical data: $L=10^{-3}$m, $W=9.2 \times 10^{-5}$ m and $W=2.03 \times 10^{-4}$ m for the narrow and broad samples, respectively, and for the thickness t we suppose 4 layers (see **section 3.1**) with 0.34 nm thick per layer, that is, $t=1.36 \times 10^{-9}$ m; which yield values for the electrical conductivity of $1.04 \times 10^6$/Ωm and $9.05 \times 10^5$/Ωm for the narrow and broad samples, respectively. These values are close to the reported $1.1 \times 10^6$/Ωm obtained by fitting the results for a variety of multilayer graphene films synthesized by CVD [15].

Regarding the linear temperature dependence of the conductivity, theoretically this dependence has been predicted for a monolayer in the ballistic regime [16] and for bilayer graphene for diffusive transport mediated by disorder [17], in both cases at high temperatures. But it should be noted that even in the early theoretical work on graphite, a linear dependence of the electrical conductivity along the graphene layers was also predicted [18]. Although bilayer graphene was grown on the copper wire in our case, in the collapsed situation and owing to its lateral dimension, the film can be considered as a four layer graphene. But some care should be taken since the lateral edges along the dimension L (see schematic in **Figure 3(a)**) may contribute to the electrical conduction because ideally the edges are curved and closed borders.

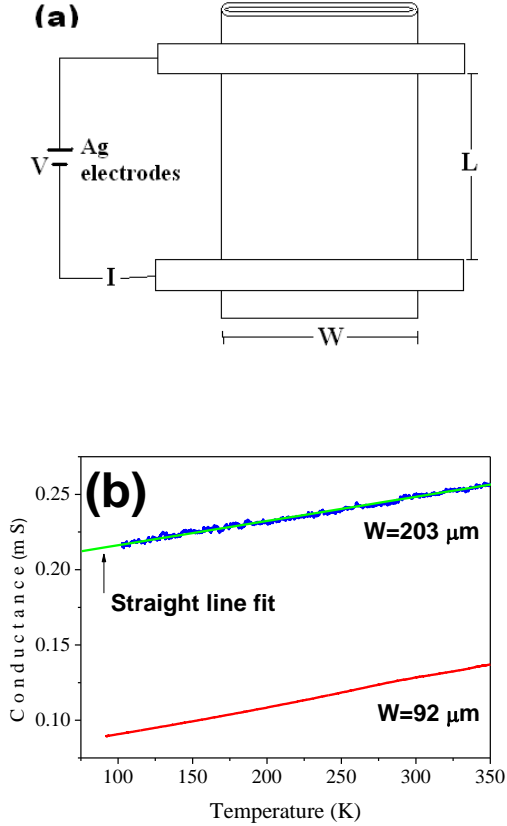

**Figure 3.** (a) Schematic of the geometry of the multilayer graphene stripe along with electrical contacts. (b) Measured electrical conductance as a function of temperature for the two kinds of films. The corresponding sheet resistance at room temperature is: ~707 Ω/sq for W=92 μm and ~812 Ω/sq for W=203 μm.

### 3.3. A Cross-Stripe Junction Device

On the other hand, taking advantage of the shape of the obtained multilayer graphene films, we built devices in the typical cross-stripe junction used in tunneling spectroscopy [19]. Firstly, two copper wires (~5mm long, 63 μm diameter) covered with multilayer graphene were crossed and fixed on a glass slide substrate using an epoxy glue in their extremities, then copper was etched and the sample washed with deionized water. Current (I) against voltage (V) characterization was made by injecting current into two adjacent arms and measuring voltage across the opposite arms of the device [19]. In **Figure 4(a)** conductance near zero bias as a function of temperature of the junction is shown, and in **Figure 4(b)** I(V) curves at three different temperatures in a log-log scale are presented. Due to the geometrical configuration of the device one should expect that the primary contribution to the electrical transport across the junction is perpendicular to the graphene layers, at least in the layers in close contact between the two stripes. An estimation of the electrical conductivity of the junction can be done as follows. The area corresponds to the intersection of the two stripes (92 μm x 92 μm in this case), for the length the graphite interlayer distance is taken as a first approximation, the conductance at room temperature is (see **Figure 4(b)**) $15 \times 10^{-3}$/Ω; all these finally yield to a value of ~$6 \times 10^{-4}$ /Ωm for the conductivity. If this value is taken as the conductivity perpendicular to the layers, and a value of ~$1 \times 10^{6}$/Ωm (see **section 3.2**) for the conductivity along the graphene layers; then an anisotropy factor of the parallel to the perpendicular conductivity of ~$1.7 \times 10^{9}$ is obtained. Clearly, this value does not represent a physical characteristic of the graphite structure because a value of ~$3 \times 10^{3}$ for the anisotropy factor for crystalline graphite has been reported [20]. In other words: the conductivity of the junction between the two stripes is at least of the order of $10^{5}$ less than the conductivity of the contact between two graphene layers in the ideal structure of graphite. Due to this fact, and that the conductance near zero bias decreases when temperature decreases (**Figure 4(a)**) and also because the differential conductance increases as the voltage bias increases (**Figure 4(b)**), it is very likely that the cross-stripe device is a tunnel junction [19].

As a last observation, it should be noted that superlinear behavior on the current dependence $I \sim V^{\alpha}$, specifically with α=3/2, has been predicted for graphene within the framework of Schwinger´s pair production and Klein tunneling [21-24]. Under some specific conditions, a linear behavior for small voltages is also found [23, 24]. As a guide for the eye, in **Figure 4(b)** the linear and superlinear ~$V^{3/2}$ behaviors are plotted. Note that the linear behavior is reasonably well reproduced for low voltages, being a small vertical shift the difference for the three temperatures, and a ~$V^{3/2}$ tendency appears to be a good option for higher voltages. We believe that our device has the appropriate geometry to observe tunneling phenomena, possibly Klein tunneling, because particles tunnel from one electrode to the other through a barrier, that in this case could be vacuum. Probably, tunneling takes place between the two adjacent parallel graphene layers of the multilayer graphene stripes that form the junction. As the voltage bias across the junction is changed, there is a relative shift of the Fermi level on the two sides of the barrier, and therefore scanning a range of energies around the Fermi energy [19] of graphene. In any case, it would be interesting to built hybrid structures, using multilayer graphene stripe as one electrode and superconducting or magnetic films as the counter electrode in the cross-stripe geometry. Experiments in this direction are currently in process in our laboratory.

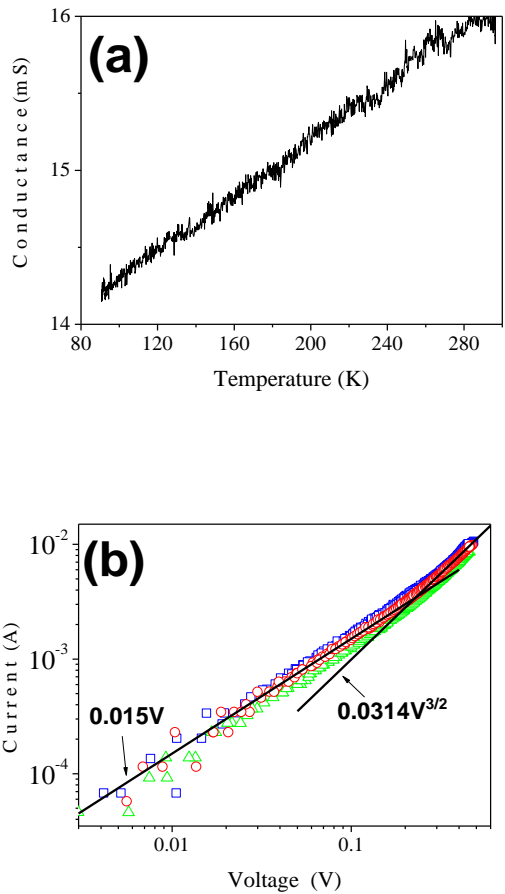

**Figure 4.** Electrical characteristics of the cross-stripe junction device. (a) Conductance against temperature measured at a fixed current of 10μA. (b) Current as a function of voltage for positive and negative polarities at different temperatures: green triangles (89 K), red circles (296 K) and blue squares (380 K).

## 4. Conclusions

Bilayer graphene was grown on copper wires by means of CVD method with methane as the carbon source. After etching the copper wire, the bilayer tube collapses forming stripes of four-layer graphene. A linear dependence of the electrical conductance as a function of temperature is found for this kind of films. Using the material obtained by this method, a cross-stripe junction is built, its electrical conductance behavior as a function of voltage bias and temperature indicates that the device is a kind of tunnel junction. It is proposed that this kind of device might be appropriate to observe Klein tunneling.

## 5. Acknowledgments
I thank Carlos Flores IIM-UNAM for the transmission electron microscopy observations.